\documentclass[preprint,preprintnumbers,amsmath,amssymb]{revtex4}
\usepackage{amsfonts}
\usepackage{mathrsfs}
\usepackage{graphicx}
\usepackage{dcolumn}
\usepackage{bm}
\begin{document}
\draft
\title{Entanglement, fidelity, and quantum phase transition in antiferromagnetic-ferromagnetic alternating Heisenberg chain}
\date{\today}
\author{Jie Ren}
\author{Shiqun Zhu\footnote{Corresponding author,E-mail: Szhu@suda.edu.cn}}

\affiliation{School of Physical Science and Technology, Suzhou
University, Suzhou, Jiangsu 215006, People's Republic of China}

\begin{abstract}

The fidelity and entanglement entropy in an
antiferromagnetic-ferromagnetic alternating Heisenberg chain are
investigated by using the method of density-matrix
renormalization-group. The effect of anisotropy on fidelity and
entanglement entropy are investigated. The relations between
fidelity, entanglement entropy and quantum phase transition are
analyzed. It is found that the quantum phase transition point can be
well characterized by both the ground-state entropy and fidelity for
large system.

\vspace{1 cm}



\end{abstract}
\maketitle

In condense matter physics, quantum phase transitions imply
fluctuations, which happened at the zero temperature~\cite{Sachdev}.
When a controlling parameter changes across critical point, some
properties of the many-body system will change dramatically. Many
results show that entanglement existed naturally in the spin chain
when the temperature is at zero. The quantum entanglement of a
many-body system has been paid much attention since the entanglement
is considered as the heart in quantum information and
computation\cite{Nielson,Bennett}. As the bipartite entanglement
measurement in a pure state, the von Neumann entropy\cite{Bennett1}
in the ferromagnetic\cite{Popkov} and antiferromagnetic
\cite{Vidal,Latorre} spin chains are investigated respectively. By
using the cross fields of quantum many-body theory and
quantum-information theory, von Neumann entropy is applied to detect
quantum critical behaviors \cite{Preskill,Osborne,Amico,Gu,Kitaev}.
A typical example is that Osborne solved exactly one-dimensional
infinite-lattice transverse-field Ising model to obtain entropy by
the Jordan-Wigner transform. The entropy predicts the quantum phase
transition point successfully\cite{Osborne}. Moreover, another
concept from quantum information science, ground state fidelity has
been used to qualify quantum phase transitions successfully in the
last few years
\cite{Zhou01,Zanardi,Oelkers,Cozzini,Venuti,Buonsante,You,Yang,Zhou02,Tzeng}.
It is shown that the fidelity and the entanglement entropy have
similar predictive power for identifying quantum phase transitions
in the most system. However, the ground state fidelity is a
model-dependent indicator for quantum phase transitions. It can not
be used to characterize the quantum phase transition in Heisenebeg
model with next-nearest-neighbor interavtion for
finite-size\cite{Chen}. Similarly, the fidelity can not detect a
BKT-like phase transition, which happens at $\Delta=1$ in
antiferromagnetic anisotropic Heisenberg model\cite{Chen}.

Recently, It is reported that $(CH_3)_2CHNH_3CuCl_3$  is realization
of the spin-1/2 alternating antiferromagnetic-ferromagnetic(AF-F)
chain by nearly the same strength of the antiferromagnetic and
ferromagnetic couplings. There are examples for the alternating AF-F
spin-1/2 chain compounds such as $[Cu(TIM)]CuCl_4$\cite{Hagiwara},
$CuNb_2O_6$\cite{Kodama}and $DMACuCl_3$ \cite{Stone}. The ground
state properties of the alternating AF-F spin-1/2 chains have been
intensely studied\cite{Hida,Kohmoto,Yamanaka}.   Entanglement,
fidelity and their relations with quantum phase transition in the
system like these materials need to be investigated further.

In this paper, the fidelity  and  entanglement entropy in the
spin-1/2 alternating AF-F chain with anisotropic interaction are
investigated. Firstly, the effect of anisotropic interaction on
ground state fidelity is investigated. Secondly, the effect of
anisotropic interaction on entropy is calculated. Thirdly, their
relations with quantum phase transition are analyzed. At last, a
discussion concludes the paper.

The Hamiltonian of  an antiferromagnetic-ferromagnetic alternating
Heisenberg chain with anisotropy of N sites is given by

\begin{widetext}
\begin{equation}
\label{eq1}H=J_{AF}\sum_{i=1}^{N/2}(S^x_{2i-1}S^x_{2i}+S^y_{2i-1}S^y_{2i}+\Delta_{AF}S^z_{2i-1}S^z_{2i})\\
             +J_{F}\sum_{i=1}^{N/2}(S^x_{2i}S^x_{2i+1}+S^y_{2i}S^y_{2i+1}+\Delta_F
             S^z_{2i}S^z_{2i+1}),
\end{equation}
\end{widetext}
where $S^{\alpha}_i(\alpha=x, y, z)$ are spin operators on the
$j$-th site, $N$ is the length of the spin chain. $J_F$ and $J_{AF}$
denote the ferromagnetic and antiferromagnetic couplings
respectively. $\Delta_{AF},\Delta_{F}$ are anisotropic interaction.
The AF-F alternating  spin chain can be regarded as the spin-1
antiferromagnetic chain in the large ferromagnetic coupling
limit\cite{Hida,Kohmoto,Yamanaka}. In the paper, open boundary
condition is considered, and we set $J_{AF}=-J_F=1,\Delta_{AF}=1$
and $\Delta_{F}>0$.

Ground state fidelity can be applied to detect the existence of the
quantum phase transitions. The definition of ground state fidelity
is shown as following. A general Hamiltonian of quantum many-body
systems can be written as $ H(\lambda)=H_0+\lambda H_I $ where $H_I$
is the driving Hamiltonian and $\lambda$ denotes its strength.
Supposes $|Gs\rangle$ represents the ground state of the system. The
ground-state fidelity between $|Gs(\lambda)\rangle$ and
$|Gs(\lambda+\delta)\rangle $ is defined as

\begin{equation}
\label{eq2}F(\lambda,\delta)=|\langle
Gs(\lambda)|Gs(\lambda+\delta)\rangle|.
\end{equation}
Because $F(\lambda,\delta)$ reaches its maximum value $F_{max}=1$
for $\delta=0$, on expanding the fidelity in powers of $\delta$, the
first derivative $\frac{\partial F(\lambda,\delta=0)}{\partial
\lambda}=0$. By using the property, the fidelity can written by

\begin{equation}
\label{eq3}F(\lambda,\delta)\simeq1+\frac{\partial^2F(\lambda,\delta)}{2\partial\lambda^2}|_{\lambda=\lambda'}\delta^2
\end{equation}
therefore, the average fidelity susceptibility $S(\lambda,\delta)$,
is given by\cite{Cozzini,Buonsante}

\begin{equation}
\label{eq4}S(\lambda,\delta)=\lim_{\delta\rightarrow
0}\frac{2[1-F(\lambda,\delta)]}{N\delta^2}.
\end{equation}

It is well known that it is hard to calculate the ground state
fidelity because of the lack of knowledge of the ground state
function. For models that are not exactly solvable, most of
researchers resort to exact diagonalization to obtain the ground
state for small size. This method can not precisely quantify the
quantum phase transition because the size of the system is too
small. The method of density-matrix renormalization-group
(DMRG)\cite{white,U} can be applied to obtain the ground state of
the model. Moreover, the technology of calculating the overlap of
two different ground states by the DMRG has been used for about a
decade\cite{Qin01,Qin02,McCulloch}. The method is used to calculate
the ground state fidelity susceptibility. We calculate $N$ up to
$78$. $\delta=0.001$ is used like in Ref.\cite{Tzeng}. The total
number of density matrix eigenstates held in system block $m=128$ in
the basis truncation procedure. The Matlab codes of DMRG  with
double precision are performed in private computer, and the
truncation error is smaller than $10^{-12}$.  The fidelity
susceptibility $S$ is plotted as a function of anisotropic
interaction $\Delta_F$ for different sizes in Fig. 1. It is shown
that one peak is exist. The maximal value increases with size
increases. The location of the maximal value deceases with size
increases. When $N=78$, the location of the maximal value
$\Delta^{max}_F=2.32$.

For comparison, ground state entanglement entropy is used to detect
the quantum phase transition point too. The definition of
entanglement entropy is given as follow. Let $|Gs\rangle$ be
assigned to the ground state of a chain of $N$ qubits, the reduced
density matrix of right-hand $L$ contiguous qubits can be written as
$\rho_L=Tr_{(N-L)}|Gs\rangle\langle Gs|$. The bipartite entanglement
between the right-hand $L$ contiguous qubits and the rest subsystem
can be measured by the entanglement entropy as

\begin{equation}
\label{eq5}E(L,N)=-Tr(\rho_L\log_2\rho_L).
\end{equation}
By using the method of density-matrix renormalization-group (DMRG),
the entropy of ground state for large system can be calculated. The
entropy of ground state is plotted as a function of anisotropic
interaction $\Delta_F$ with sizes of $N=20,40,60,80,100$ with in
Fig. 2. The largest states of $m=64$ is kept and $L=N/2-1$ is
considered. There is one peak too. The peak $E_{max}$ increases with
the size increases. The corresponding critical point
$\Delta^{max}_F$ of the peak decreases with the size increases.

As we known, the peak in fidelity susceptibility and the peak in
entanglement entropy indicate the quantum phase transition. To be
more precise, we also investigate whether the positions of the
extreme points of entanglement entropy and fidelity in an infinite
system. The results for the scaling of entanglement\cite{Latorre1}
and fidelity\cite{Gu02} can be used to investigate the quantum phase
transition point. We plot the maximum entropy of entanglement and
fidelity susceptibility as a function of the inverse size of the
system. A numeric fit is made, the results is shown in Fig. 3. They
apparently agree with linear scaling. In the thermodynamic limits,
their values tend to $\Delta^{c}_F=2.3$, respectively. Both the
extreme points of fidelity susceptibility labeled by $S$ and entropy
labeled by $E$ represent indeed the quantum phase transition. As we
known, when $\Delta_F=1$, the phase is the Haldane
phase\cite{Hida,Kohmoto,Yamanaka}. In the limits
$\Delta_F\rightarrow\infty$, the phase is Ising model universality
class state. Our result is similar with the result
$\Delta_{AF}/2\simeq J_F/J_{AF}$\cite{Yamanaka}. It confirms further
that the Haldane-Ising transition occurs at the point.

In the paper, the fidelity susceptibility and entropy in the
Heisenberg chain with the Alternating AF-F interaction are studied.
By the present DMRG calculations for the model, the effect of
anisotropic interaction on fidelity susceptibility and entropy in
large size is presented. Their relations with quantum phase
transition are investigated. It is shown that the point of the
quantum phase transition is clearly marked by the peak of the
fidelity susceptibility and entropy. The critical point in the
thermodynamic limit is obtained by using the finite-size scaling
theory. The fidelity susceptibility and the entanglement entropy can
have similar predictive power for revealing quantum phase transition
in the system.

\vskip 0.4 cm {\bf Acknowledgments}

It is a pleasure to thank Yinsheng Ling and Jianxing Fang for their
many helpful discussions. The financial supports from the
Specialized Research Fund for the Doctoral Program of Higher
Education of China (Grant No. 20050285002) and the National Natural
Science Foundation of China (Grant No. 10774108) are gratefully
acknowledged.

\newpage
\begin{figure}
\includegraphics{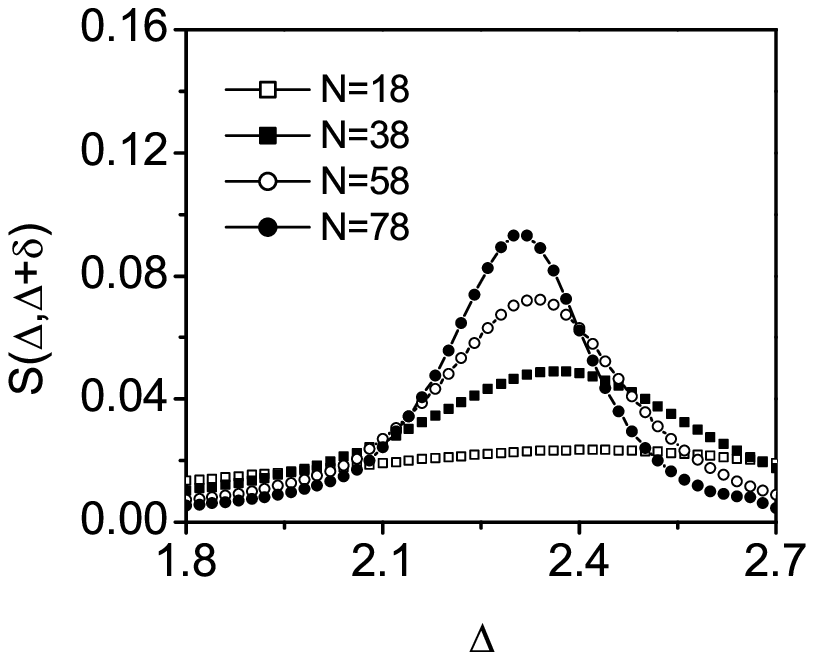}\caption{The fidelity suscepility $S$  is plotted as a
function of anisotropic interaction $\Delta_F$ for different sizes.
}
\end{figure}

\clearpage
\newpage
\begin{figure}
\includegraphics{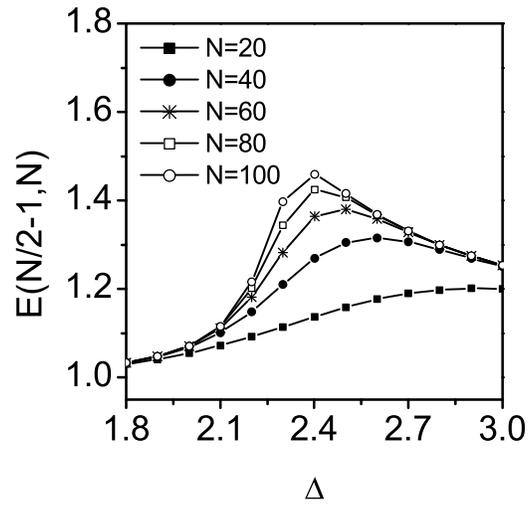}\caption{The entropy $E$  is plotted as a function of
anisotropic interaction $\Delta_F$ for different sizes. }
\end{figure}

\clearpage
\newpage

\begin{figure}
\includegraphics{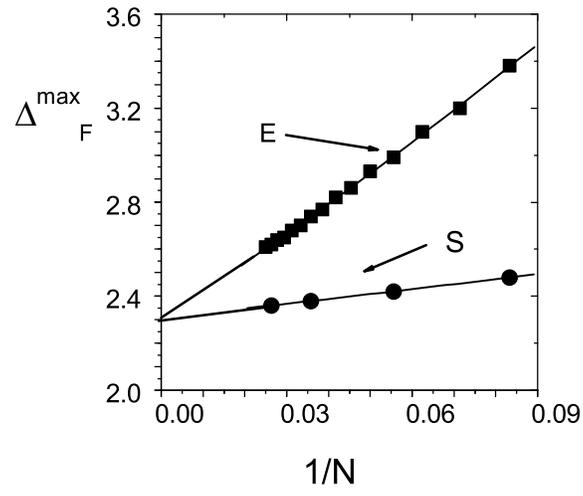}\caption{Scaling of $\Delta_F^{max}$ versus $N^{-1}$. The $\blacksquare$ for
entropy and $\bullet$ for fidelity susceptibility are obtained by
numeric stimulation and the lines are the fit lines.}
\end{figure}

\end{document}